\documentclass[fleqn,12pt]{wlscirep}
\usepackage[utf8]{inputenc}
\usepackage[T1]{fontenc}
\usepackage[version=4]{ mhchem }
\usepackage{graphicx}%
\usepackage{siunitx}
\usepackage{multirow}%
\usepackage{amsmath,amssymb,amsfonts}
\usepackage{amsthm}%
\usepackage{mathrsfs}%
\usepackage[title]{appendix}%
\usepackage{xcolor}%
\usepackage{textcomp}%
\usepackage{manyfoot}%
\usepackage{booktabs}%
\usepackage{algorithm}%
\usepackage{algorithmicx}%
\usepackage{braket}
\usepackage{algpseudocode}%
\usepackage{listings}%
\usepackage{mhchem} 
\usepackage{lineno}
\usepackage{makecell}
\usepackage{float}
\usepackage{tabularx}
\usepackage[numbers]{natbib}

\title{ The Gamma Factory: \\ 
A New Experimental Paradigm for CERN's HL-LHC--FCC-ee Transition}

\vspace{10 mm}

\author[1,2,*]{Mieczyslaw Witold Krasny}

\affil[1]{LPNHE, IN2P3, CNRS, University Paris Sorbonne, Paris, 75005, France}

\affil[2]{CERN,  Geneva, 1211, Switzerland}

\affil[*]{email: mieczyslaw.witold.krasny@cern.ch}

\begin{abstract}

The Gamma Factory (GF) proposal \cite{Krasny:2015ffb} is motivated by the recognition of a largely untapped potential of the CERN accelerator complex to enable a new research programme at the intersection of particle, nuclear, atomic, fundamental, and applied physics. These fields could benefit from novel experimental tools made possible by a future GF facility.
The central concept is to produce, accelerate, cool, and store atomic beams of highly relativistic partially stripped ions in the LHC, which would serve as an effective atomic trap. The internal degrees of freedom of these ions are then resonantly excited using laser photons. In the GF scheme, laser-cooled atomic beams serve both as high-precision probes and as low-emittance beam sources for high-luminosity LHC operation in the ion-ion collision mode.
Interactions between laser pulses stored in Fabry--Perot cavities and circulating ion beams give rise to high-energy, highly collimated, and polarised secondary photon beams. Their expected intensities exceed those of existing gamma-ray sources by several orders of magnitude. These photon beams can further be used to generate unprecedented-intensity, tertiary beams of polarised electrons, positrons, muons, neutrons, radioactive ions, and flavour- or CP-tagged neutrinos.
Furthermore, under a specific configuration, the same photon-driven processes may be exploited in an energy-production scheme generating the requisite plug-power for LHC operation.
Together, cold relativistic atomic beams, high-intensity photon beams, and tertiary beams constitute a versatile experimental platform capable of opening a wide range of new scientific opportunities at CERN. By exploiting existing accelerator infrastructure and available state-of-the-art laser technologies, the GF offers a path to a cost-effective and timely programme capable of sustaining experimental innovation and bridging the gap between the HL-LHC era and the future FCC era.

\end{abstract}
\begin{document}

\flushbottom
\maketitle
\thispagestyle{empty}

\section*{Introduction}

\vspace{5 mm}
{\it "Life is like riding a bicycle. To keep balance, you must keep moving." -- A. Einstein}
\vspace{5 mm}

The history of science shows that every discipline, after 
a period of rapid expansion, eventually enters a mature 
phase in which progress slows. Research becomes focused on
revisiting well-established questions and achieving only 
incremental improvements in precision and scope. Particle 
physics is no exception to this rule. Following the 
extraordinarily creative decades that led to the construction of the 
Standard Model of elementary particles and their 
interactions, the field has now reached such a phase.

Three principal strategies have driven 
particle physics research over the last three decades:
\begin{itemize}
\item precision tests of the Standard Model with available particle beams,  
\item dedicated searches for new particles and phenomena 
       predicted by beyond-the-Standard-Model theories,  
\item construction of new accelerators and colliders, and development of novel research tools.
\end{itemize}

Today, the first two strategies are becoming 
progressively less attractive. Despite continuing—though now saturating—advances in experimental precision, 
the Standard Model remains unchallenged. Furthermore, no 
robust theoretical predictions identify new particles or 
phenomena that could be accessible with present accelerator 
and detector technologies, or their incremental upgrades.

The third strategy now offers the most promising route to 
reinvigorating particle physics, in particular if the 
following two conditions are satisfied:
\begin{enumerate}
\item 
Technological leaps—rather than incremental 
improvements—are achieved, enabling substantial 
increases in particle-beam intensity, energy, or 
collision luminosity.
\item 
The cost of the required new research infrastructures remains affordable.
\end{enumerate}

Several options that attempt to satisfy both conditions are currently under discussion.
These include the construction of colliders based on established technologies, 
such as the FCC-ee \cite{FCC:2018byv,FCC:2018evy,FCC:2018vvp}; 
the development of novel acceleration or collision techniques for
conventional beams \cite{Foster:2023bmq}; and the construction of entirely new types of colliders in which 
electrons, protons, or ions are replaced by short-lived 
muons \cite{InternationalMuonCollider:2024jyv}.
The option that has presently received the strongest support from the particle physics community 
is the FCC-ee collider \cite{FCC:2018byv,FCC:2018evy,FCC:2018vvp},
which would extend the energy frontier of electron--positron collisions by a factor of up to 1.7 and the luminosity frontier by two to four orders of magnitude. 

Pursuing the high-energy frontier research  is well motivated, in particular, to improve the precision of canonical 
Standard Model measurements. However, it cannot be taken for granted  
that any of the presently discussed high-energy frontier 
colliders will achieve a sufficient increase in collision energy 
to enable the discovery of new particles or phenomena. 
This represents a qualitatively new situation, contrasting 
sharply with earlier generations of energy-frontier 
colliders, such as the S$p\bar{p}$S and the LHC at CERN, or
the Tevatron at Fermilab, where the discoveries of the W 
and Z bosons, the Higgs boson, and the top quark were 
strongly anticipated based on precision measurements at lower 
energies interpreted within the framework of quantum field 
theory.

What can be stated with confidence is that none of the 
proposed next-generation high-energy frontier  
colliders will be operational before the current research programmes at the 
world's major accelerator centres reach their precision and discovery potential saturation. This creates an urgent need for 
new ideas and complementary strategies that can provide genuine 
technological breakthroughs on shorter timescales, 
ensuring continuity in the exploratory, discovery-driven branch of particle physics research,  
regardless of whether—or when—a future energy-frontier 
collider becomes operational.

Strategies aimed at bridging the period between the end of the HL-LHC programme and the FCC-ee era at CERN, and requiring the development of novel research tools, were not addressed in the recent European Strategy Update for Particle Physics, either in its discussions or in its final recommendation document \cite{ESPPU2026}. 

The motivation for this paper is to present a project developed over the last decade that could become an essential component of such a strategy. 

The Gamma Factory (GF) project, discussed in this paper, 
has three key merits: (1) it complements high-energy frontier studies with a high-intensity frontier programme; (2) it can fill the gap between 
the end of the HL-LHC era and the FCC-ee era with a rich, multidisciplinary research programme; and (3) its implementation cost is sufficiently low that it does not significantly affect the financing of future high-energy frontier collider construction. 

The GF project \cite{Krasny:2015ffb} was motivated by the recognition that substantial technological advances can 
be achieved by using the existing CERN accelerator infrastructure
in unconventional ways and by exploiting rapid 
developments in laser science and technology that have 
already transformed many other fields.

The aim of the GF is to significantly extend CERN’s 
scientific programme across particle, nuclear, atomic,
fundamental, and applied physics. It seeks to deliver technological advances that open new 
research opportunities and create the potential for 
unexpected discoveries, all at relatively modest economic, 
ecological, and personnel cost.

It is important to emphasise that the primary goal of the
Gamma Factory is to develop new research tools for the 
high-intensity and high-precision frontier, rather than 
to execute a predefined programme driven by specific 
beyond-the-Standard-Model scenarios.

Historically, new directions in science have more often 
been driven by the development of new tools than by new theoretical
concepts. Theory-driven revolutions reinterpret known 
phenomena, whereas tool-driven revolutions reveal new ones. The 
Gamma Factory is conceived in this spirit.

\section{Gamma Factory in a  Nutshell}

\vspace{5 mm}
{\it "Nothing is truly real  but atoms and void..." -- Democritus}
\vspace{5 mm}

The key requirement for the Gamma Factory research 
programme is to create a new type of beams--the 
ultra-relativistic {\bf atomic beams} of highly-charged 
Partially Stripped Ions (PSI)--and store them in the 
existing CERN storage rings. 
The CERN storage rings are proposed to play the role of  giant 
atomic traps, in which the atomic or nuclear degrees of 
freedom of the beam particles are resonantly excited by 
the laser light.

The already existing CERN accelerator infrastructure is 
capable \cite{Hirlaender:IPAC18-THPMF015,Kroger:2019wfh,Schaumann:2019evk,Gorzawski:2020dgx} to create, accelerate and store such 
beams 
at very high energies, over a large range of the Lorentz
factor: 
$ 200 < \gamma_L < 3000$, at  high bunch intensities: 
$ 10^8 < N_{\rm bunch} <5 \times 10^9 $, and at a  bunch 
repetition rate of up to $20$~MHz.

Gamma Factory aims to exploit the recent, impressive progress in laser technology 
and incorporate its achievements into the particle-accelerator-based 
research \cite{Roikova:2025kfu,Granados:2024sht}. 
Laser photons enable 
efficient manipulation of stored PSI beams by the 
selective resonant excitation of the beam particles.
The GF  scheme provides a unique possibility for high-precision, experimental 
investigation of the Schwinger limit of strong QED processes in high-Z atomic 
systems\cite{Budker:2020zer},  and contributes to the  
development of nuclear clocks \cite{Jin:2022tpv}.
Moreover, selective atomic-level excitations of PSIs  
can  enrich the LHC collider research  
by employing  highly efficient GF 
methods of longitudinal and transverse beam cooling of the high-energy beams, 
which employ and extend the techniques developed and mastered in atomic 
physics research \cite{Krasny:2020wgx,Krasny:2021llv}. 
Additionally, PSI beams can be used as the source of 
unprecedented-intensity secondary photon beams. 

The unique opportunity, existing presently only at CERN, 
to conduct the Gamma Factory research programme 
comes from the appropriate matching of the available 
energy range of the  PSI beams with the wavelength range 
of available lasers. Resonant excitation of atomic levels
by laser photons becomes possible thanks to high 
energies of the PSIs stored at the CERN's storage rings. 
For the first time, profiting from  the large 
relativistic Lorentz factor $\gamma _{L}$ of the LHC or SPS PSI 
beams, all the atomic degrees of freedom, including those of 
highest-$Z$ atoms, can be resonantly excited  by the infrared, 
visible, UV or  EUV 
laser photons, by exploiting  the Doppler upshift of  
the laser-photon energies  by a factor of $2\gamma _{L}$.

\begin{figure}[!htpb]\centering
	\includegraphics[width=0.90\linewidth]{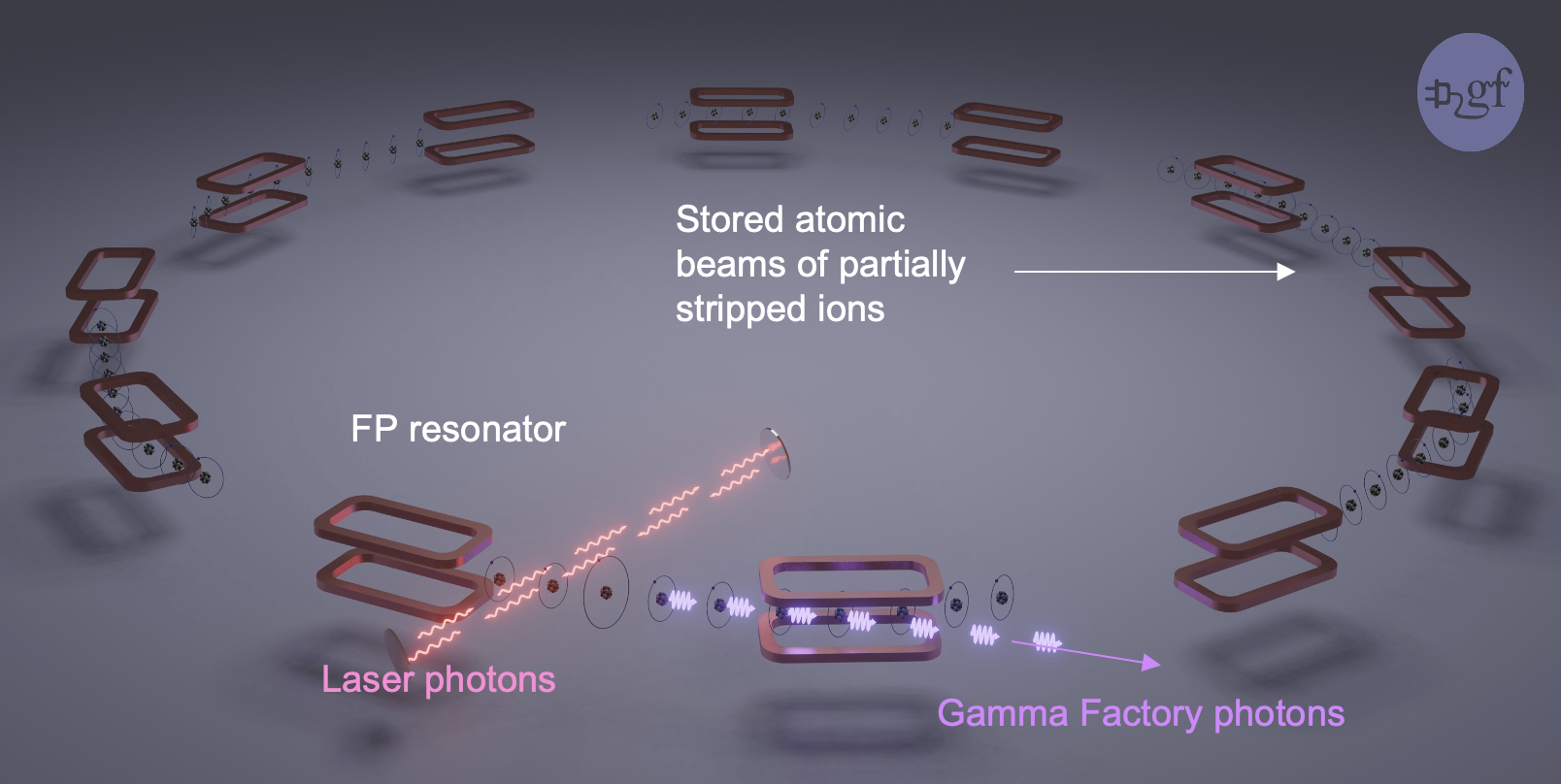}
	\caption{The GF photon source scheme. For its  high-intensity implementation, the Fabry-Perot (FP) cavity is used to recirculate laser pulses.   
	 }
    \label{Fig:GF_photon_source}
\end{figure}

The laser-photon absorption process is followed by the spontaneous photon emission.
Photons emitted in the direction of the ion beam have their energy Doppler-boosted by
a further factor of $2 \gamma _L$. As a consequence,  the combined process of absorption
and re-emission boosts the photon energy by a factor  of up to $4 \gamma _L ^2$ 
-- a factor reaching the $\sim 10^8$ value for the LHC  beams stored at their 
maximal energy.  GF scheme thus allows to convert the eV-scale laser 
photons to keV-, or MeV-energy-range photons. Their narrow energy band can be 
precisely selected,  over the broad energy region ranging between 40 keV and 400 MeV,  
through a research-goal-specific choice of the atomic 
number of the PSI, the number of electrons it carries,  and by the concrete choice of
the laser type. 
The principle of the GF  photon source is illustrated 
in Fig.\,\ref{Fig:GF_photon_source}.

\begin{figure}
    \centering
    \includegraphics[width=0.95\linewidth]{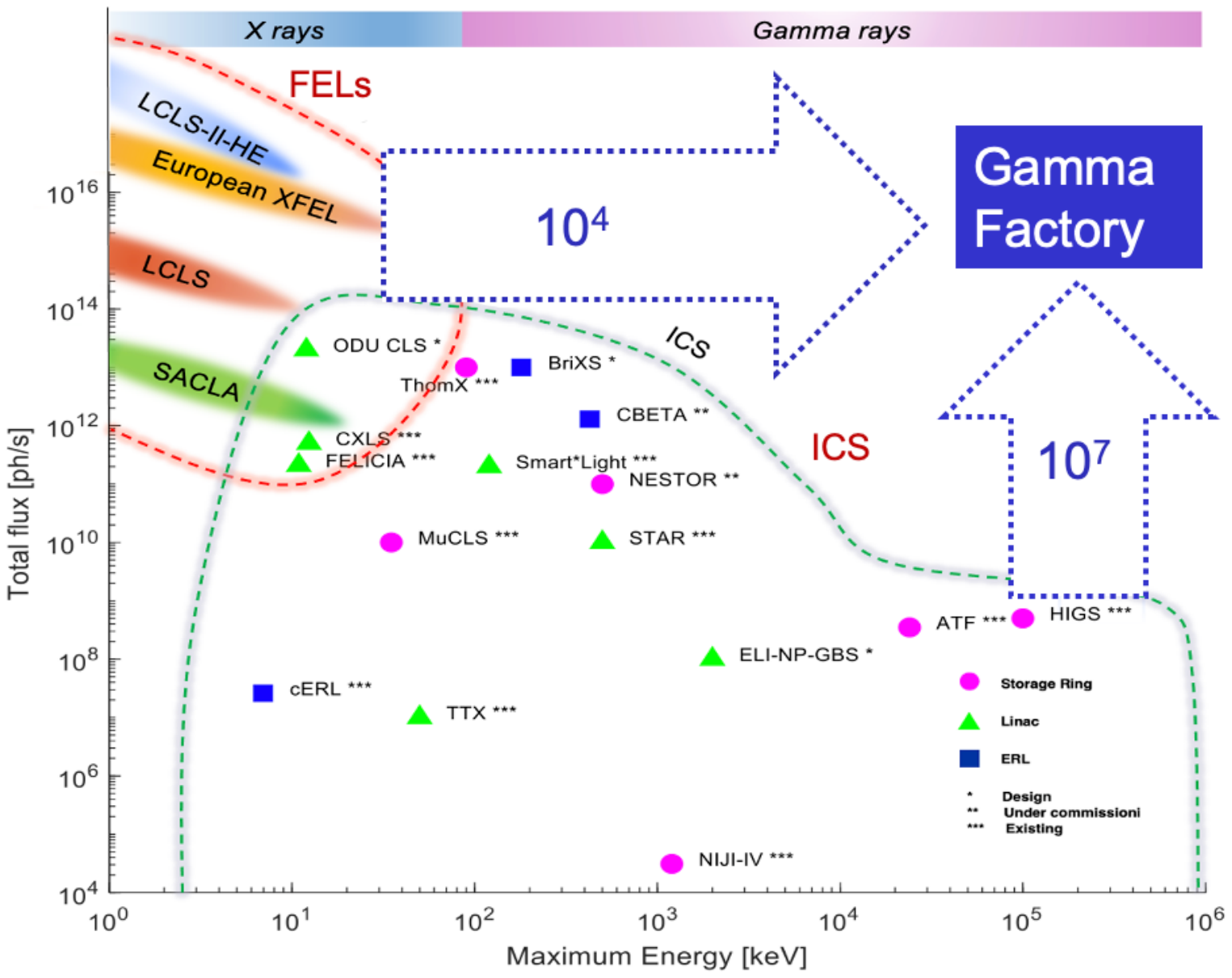}
    \caption{The expected energy and the photon flux of the GF photon source, compared to the existing, and the future Free Electron Laser (FEL), and Inverse Compton Scattering (ICS) photon sources. Gamma Factory aims to extend the FEL's photon energy range by a factor of 10$^4$, while preserving FEL-like fluxes. In the ICS energy region,  a leap in the achievable photon flux,  by a factor of up to  10$^7$, is expected in the GF scheme. }
    \label{Fig:gamma_factory_energy_flux}
\end{figure}

Since the resonant-peak absorption cross-section is up to  nine orders of magnitude larger than  
that for the photon-electron collisions, a leap of many orders of magnitude 
in the photon source intensity is expected. The 
expected leap in the GF photon source flux and energy with respect to the existing and the future electron beam-based sources is 
presented  in 
Fig.\,\ref{Fig:gamma_factory_energy_flux}.

Thanks to the large $\gamma _{L}$ of the CERN PSI beams 
 the emitted photons appear (in the laboratory frame) to form a highly collimated 
$X$-ray or $\gamma$-ray beam, which can be extracted from the 
collision zone and used directly,  or indirectly -- for producing unprecedented-intensity tertiary beams of neutrons, pions, muons, neutrinos, and radioactive ions.

In summary, the key GF idea, essential for creating {\bf novel tools} for the 
high-intensity frontier research--is to exploit {\bf resonant photon 
scattering} over a wide energy range.  
Laser photons in the eV energy range excite atomic 
resonances and produce secondary,  energy-boosted photons. Energy-boosted, 
MeV-range photons excite: (1) nuclear fission resonances and produce tertiary radioactive ions, or  (2) Giant Dipole Resonances (GDRs) to produce tertiary neutrons,  or (3) the nucleon $\Delta$ resonances to produce monochromatic,  tertiary pions. The GF beam-cooling methods enhance the luminosity of the ion-beam–driven photon–photon colliders and enable experimental studies of resonant Higgs production in photon-photon collisions. 

In addition,  the GF  project can deliver a "no-cost"
monochromatic electron beam: (1)  to study the electron-proton collisions 
at the LHC \cite{Krasny:2004ue}, and (2) for its  acceleration using ion-bunch-driven plasma wakefields \cite{AWAKE:2022aeo,Cooke:2020arc}.

\section{ Gamma Factory beams and their collisions}

\vspace{5 mm}
{\it "A good tool improves the way you work. A great tool improves the way  you think." -- J. Duntemann}
 \vspace{5 mm}
 
 The current CERN research programme relies on two principal pillars:
\begin{enumerate}
    \item 
The injector-complex, extracted-beam programme, which uses the primary and secondary beams delivered by the PS, Booster, and SPS accelerators.
     \item      
The LHC collider programme, which studies proton-proton and ion-ion collisions at the highest achievable energies.
\end{enumerate}

The CERN proton injector complex underwent a major upgrade during 2019--2020, reaching its maximal performance. 
It is now capable of delivering approximately 
10$^{19}$ SPS protons per year for the North Hall fixed-target experimental programme, about 
10$^{20}$  Proton-Booster protons per year for the ISOLDE 
fixed-target programme,  
3 $\times$ 10$^{19}$ PS protons per year for the nTOF fixed-target
research programme, and up to roughly $ 3 \times$ 10$^{12}$ antiprotons per year
for the Extra Low Energy Antiproton ring (ELENA) research programme 
\cite{PBC:2025sny}.

If the current CERN fixed-target programme were extended, 
in the future to include
proton beams extracted from the LHC, the yearly integrated number of
the LHC top-energy extracted protons would be significantly smaller
than the numbers quoted above because of the long 
bunch-collection and acceleration cycle of the LHC.
No technological breakthrough is thus  expected for the 
CERN fixed target programme in the future -- even if the LHC-ring extracted 
beams were added to the  CERN's extracted beam inventory. 

How, then, can a leap of several orders of magnitude 
in the intensity of CERN's extracted beams be achieved?

The Gamma Factory project proposes a novel method for producing and
extracting beams from the LHC storage ring \cite{Krasny:2015ffb}. 
In the Gamma Factory
scheme, the stored PSI beam particles play a passive role
of the photon frequency converters and highly-efficient converters of the LHC 
storage ring RF-power into the power of the produced beams 
(see discussion in \cite{Baolong:2024ataN} for more details). 
The overall energy efficiency of such a scheme is by a large factor
better than that of  the electron-beam-driven photon sources -- 
for example,  by a factor of $\approx$ 300  better than the overall energy efficiency of the DESY FEL source \cite{Krasny_EPS_Marseille}.

In the GF scheme,  each of the stored beam particles can be reused tens of 
millions of times before it is lost and produces, over its lifetime,  tens of 
millions of photons. 
This is because the beam particle energy transmitted to the photon, of up to 400 MeV, is appreciably  smaller than the longitudinal momentum dispersion of 
the stored PSI beam particles\footnote{For the Hydrogen-like lead beam stored at the top LHC energy the longitudinal momentum dispersion of the beam particles is $\approx$ 60000 MeV.}.  
The overall beam particle loss can thus 
be compensated by the LHC RF system on 
a  turn-by-turn basis\footnote{Note, that for radial, small-lifetime excitations of high-Z ions, and for the laser pulse length 
being significantly longer than the excited-state lifetime,  each of the stored ions can produce multiple photons per turn.}.
This new concept enables the delivery of  up to
10$^{25}$ high-energy unpolarised, polarised, or twisted photons 
per year \cite{Baolong:2024ataN} for the  LHC-extracted  photon beam 
fixed-target programme,  and for the production of unprecedented intensity
tertiary beams. The GF beam extraction scheme  could increase by
several orders of magnitude,  the present intensity of the neutron
beams delivered for the nTOF experimental programme \cite{Baolong:2024ataN}, and the present
intensity of radioactive ion beams for the ISOLDE experimental
programme \cite{Nichita:2021iwa}. In addition, it enables  the production of 
unprecedented-intensity and emittance tertiary beams of polarised positrons, muons, and CP- and flavour-tagged neutrinos \cite{Apyan:2022ysh}.

The ultimate instantaneous luminosity of proton-proton 
collisions will be achieved
in the forthcoming HL-LHC phase of the LHC collider operation.
An increase in the luminosity collected so far, by a 
factor of about six is expected in this phase 
\cite{ZurbanoFernandez:2020cco}. Such an increase will  
improve the statistical precision of the LHC canonical measurements.
It is, however, unlikely to lead to discoveries of new phenomena.  The LHC 
operation mode,  for which a more significant increase in collected 
luminosity is still technologically feasible, as demonstrated by the  
BNL ion collision programme \cite{Fischer:2016jsu},  is the ion-ion
collision operation mode.  A leap in the luminosity 
of colliding ion beams--particularly interesting for colliding isoscalar-ion 
beams, for  which the weak-isospin symmetry is restored--is
required to open the unexplored domain of precision electroweak and 
Higgs-sector measurements in the ion-ion collisions. 
The electroweak research  program with high-intensity ion beams could complement, and improve 
\cite{Krasny:2020wgx,Placzek:2024bkv,Krasny:2005cb,Krasny:2007cy,Fayette:2009yvt,Krasny:2010vd}, 
the measurement precision attainable in proton 
collisions\footnote{It is worth reminding here that the most precise
deep-inelastic electron, muon, and neutrino scattering experiments--pivotal for the construction of the Standard Model--used 
isoscalar-nucleus targets rather than proton targets.}. At the same 
nucleon-nucleon luminosity,  ion-ion collisions produce smaller
number of particles within the acceptance region of the LHC-experiment's trackers, 
thus reducing the pile-up strain and the necessity of proton-proton luminosity levelling.
The high-luminosity ion-ion collisions can provide a unique 
possibility of observing Higgs-boson production in 
photon-photon collisions at 
the LHC.   
Gamma Factory   can play an important role in achieving a
significant increase in the ion collision luminosity at the LHC 
by providing a new and highly efficient cooling method for the CERN's 
high-energy ion beams\cite{Krasny:2020wgx,Kruyt:2024sty}. 

Besides the contribution to the two principal CERN research programme
pillars, Gamma Factory could open new domains of atomic physics research,
so far unaddressed at CERN, with the highly-charged Hydrogen-, Helium-, and  Lithium-like beams \cite{Zolotorev1997,Budker:2022kwg,Serbo:2021cps,Yerokhin:2022ebp}.
Such beams can be controlled at the LHC with the precision unattainable at  any other colliders, by placing the beam-control Fabry-Perot cavity in the 
dipole magnetic field, and by employing the Stark and Zeeman splitting of the quantum-states of the beam particles
(more details on such a new scheme can be found in \cite{Richter:2025nef}).  

Finally,  GF  can deliver a monochromatic electron beam 
for the LHC experimental programme. 
In the GF scheme \cite{Krasny:2004ue}, electrons are accelerated 
and stored in the LHC rings by attaching them to their carrier ions. 
Since the range of strong interactions is significantly smaller than the average distance between the electron and the nucleus, electrons can collide at the present LHC interaction points with the counter-propagating protons or fully-stripped light ions,  and remain 
unperturbed by the presence of their  (ion) carriers. The current  LHC
experiments could select and analyse such collisions without the necessity of
a hardware upgrade. 

The electrons can also be separated  from the parent ion in the stripper foil  and re-accelerated using the ion-bunch-driven plasma wakefields \cite{AWAKE:2022aeo, Cooke:2020arc}.

\begin{figure}[!htpb]\centering
\includegraphics[width=0.90\linewidth]{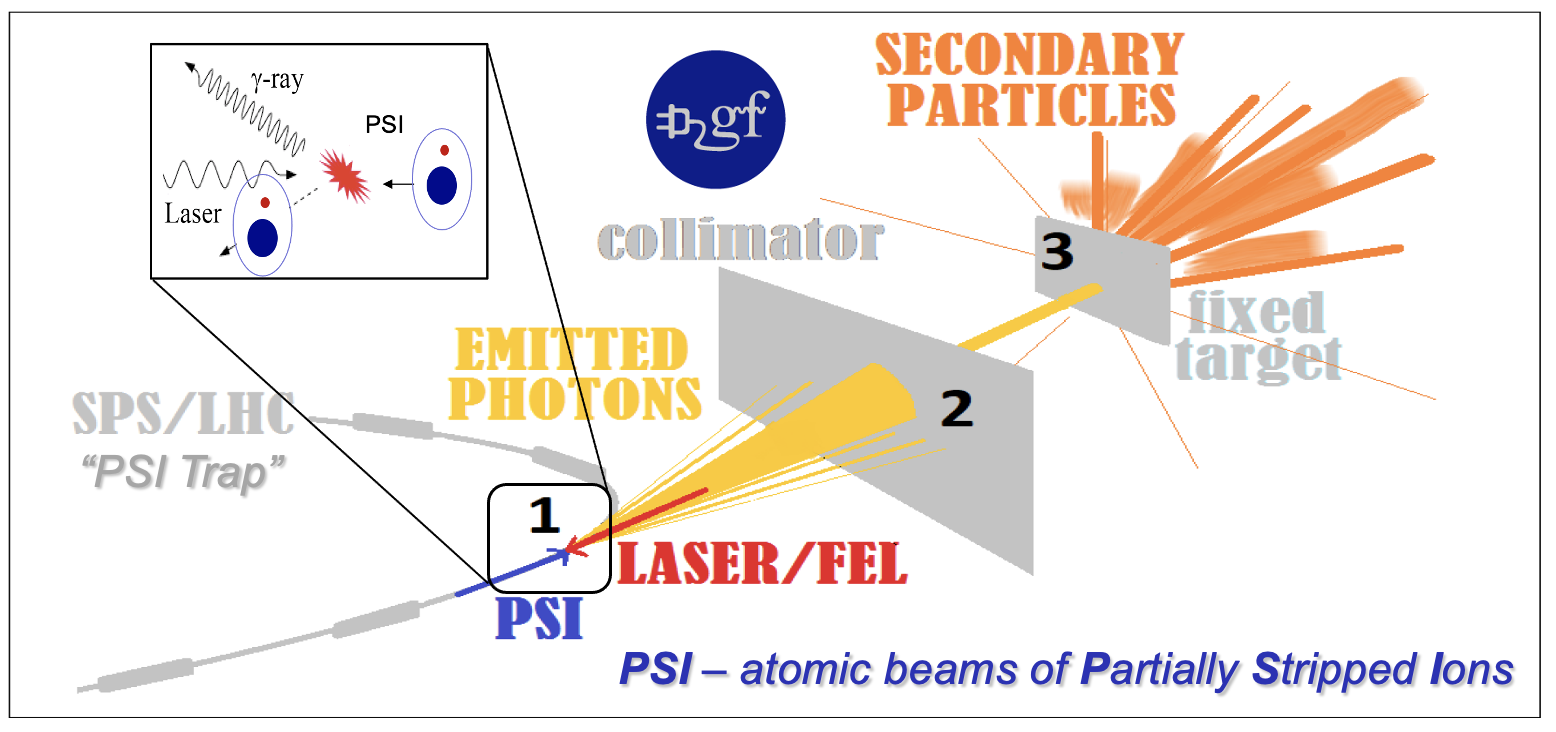}

	\caption{Primary PSI beams in the SPS/LHC storage rings and the production scheme of the GF secondary and tertiary beams.  
	 }
    \label{Fig:GF_nut_shell}
\end{figure}

Fig.\,\ref{Fig:GF_nut_shell} illustrates schematically the production scheme of the GF secondary photon beams and tertiary beams for the possible future CERN LHC-based fixed target research programme. 

\section{Gamma Factory research opportunities}

\vspace{5 mm}
{\it " The existence of the Standard Model does not imply the existence of a standardised anticipation of the future. The only thing that deserves institutionalisation is doubt. This problem of maintaining diversity of approach afflicts both experiment and theory, and if I have any concern about how the field is developing, it is about this point I worry the most." -- J.D. Bjorken (Bj)}
 \vspace{5 mm}

The Gamma Factory (GF) has the potential to extend the CERN scientific programme far beyond its present boundaries by introducing a new class of high-intensity, high-precision research tools. Rather than targeting a single physics question, it enables a qualitatively new experimental regime—combining ultra-relativistic beams, resonant photon interactions, and unprecedented control of atomic and nuclear systems. This capability opens a broad landscape of research opportunities across multiple domains of fundamental and applied science.

In particle physics, the GF programme enables a shift toward precision studies at the intensity frontier. It provides access to high-sensitivity tests of electroweak interactions, and novel studies of the Higgs boson in $\gamma\gamma$ collisions \cite{Krasny:2020wgx,Placzek:2024bkv}. At the same time, it offers new avenues for probing rare processes, such as suppressed muon decays and CP violation in the neutrino sector \cite{Apyan:2022ysh}, as well as opportunities to revisit QCD confinement and to perform electron--proton collision studies at the LHC within the present experimental framework
\cite{Krasny:2004ue}.

In nuclear and atomic physics, the GF enables precision studies that are currently inaccessible. It provides access to the strong-field QED regime approaching the Schwinger limit \cite{Budker:2020zer}. It allows nuclear spectroscopy with tunable, monochromatic, unpolarised, polarised, or twisted photon beams, detailed investigations of the interplay between atomic and nuclear degrees of freedom, and measurements of neutron distributions in nuclei \cite{Budker:2021fts,Nichita:2021iwa,Budker:2022kwg,Viatkina2019NeutronSkins}. At the atomic level, highly charged ions can be manipulated and studied with a degree of control unattainable in existing facilities, opening the way to precision tests of bound-state QED and to the exploration of exotic systems such as muonic and pionic atoms \cite{Budker:2020zer,Bieron:2021ojp,Serbo:2021cps,Flambaum:2020bqi,Richter:2025nef}.

More broadly, the GF provides new tools for fundamental physics. Its capabilities support searches for dark matter, tests of fundamental symmetries( for example,  a precision measurement of the parity violation in atomic systems\cite{Viatkina:2026xft}), and the development of novel precision instruments such as nuclear clocks and atomic interferometers \cite{Wojtsekhowski:2021xlh,Karbstein:2021otv,Balkin:2021jdr,Chakraborti:2021hfm,Jin:2022tpv,Thirolf:2019ocm,Schmirander:2024ovh,Richter2022}. These directions benefit directly from the combination of high intensity, narrow bandwidth, and controllable quantum states of the beam particles.

From the perspective of accelerator science, the GF represents a platform for transformative technological developments. It enables laser-based beam cooling at high energies, the production of low-emittance hadronic beams, and the generation of high-intensity tertiary beams of positrons, muons, neutrons, and radioactive ions \cite{Krasny:2021llv,Krasny:2020wgx,Zimmermann:2022xbv,Cooke:2020arc,Zimmermann:2022svn,Apyan:2022ysh,Zimmermann:2023rqo,Granados:2024sht}. In addition, it offers unique possibilities for producing narrow-band, CP- and flavour-tagged neutrino beams \cite{Apyan:2022ysh}, enabling  new experimental strategies in neutrino physics.

The potential impact of the GF extends beyond fundamental research. Its high-intensity, tunable photon beams create opportunities in applied physics, including energy-related technologies \cite{Baolong:2024ataN}, fusion and fission studies, production of medical isotopes and isomers, and precision lithography. 

Despite this broad potential, only a limited subset of GF applications has so far been explored in detail. Existing studies have demonstrated the feasibility of optimising ion species and charge states, laser parameters, beam dynamics, and interaction-region optics. Several practical challenges,  such as  as heat disipation  in targets, exposed to MW-class photon beams, have also been addressed.
The  results of the studies provide a solid foundation for further development, while leaving significant room for expansion into unexplored domains.

Over the past decade, the GF concept and its applications have been presented and discussed at major accelerator laboratories, including CERN, FNAL, BNL, Thomas Jefferson National Accelerator Facility (TJNAF), PSI, and DESY, and at numerous international conferences. 
Studies of the GF research opportunities have not yet entered a full development phase and currently remain in a state of limited activity, pending broader insitutional recognition of the GF's  scientific and technological potential by CERN and national funding agencies.

After several decades of primarily theory-driven searches for new particles and phenomena beyond the Standard Model, the Gamma Factory programme advocates a complementary approach: advancing the intensity frontier to study known systems—atoms and nuclei—with unprecedented precision and control. In doing so, it opens the possibility of discovering new physics through small, previously inaccessible deviations from Standard Model predictions, including effects that may lie outside the scope of existing theoretical frameworks.

\section{Feasibility}

\vspace{5 mm}
{\it "Divide each difficulty into as many parts as is feasible and necessary to resolve it." -- R. Descartes}
 \vspace{5 mm}

Over the past decade, extensive studies have been carried out to assess the feasibility of the Gamma Factory (GF) concept and to validate its key components. These studies demonstrate that the core elements of the programme are grounded in existing or already demonstrated technologies, while clearly identifying the remaining steps necessary for full implementation.

The production, acceleration, storage, and control of partially stripped ion (PSI) beams in the SPS and LHC have been successfully demonstrated \cite{Hirlaender:IPAC18-THPMF015,Kroger:2019wfh,Dutheil:2020ekk,Gorzawski:2020dgx,Schaumann:2019evk,Rebeca:2022}. In parallel, the optical system required for the GF scheme has been designed \cite{Martens:2022ptn}, and stable, high-power operation of Fabry--Perot cavities—reaching record average stored powers of 700~kW—has been achieved \cite{Lu:2024gwe,Granados:2024sht,Lu:24}. 

Dedicated SPS beam tests have demonstrated the required precision of beam position and momentum steering at the interaction point between laser pulses and PSI bunches \cite{Rebeca:2022}. In addition, a comprehensive software framework has been developed and benchmarked to simulate PSI beam dynamics, photon production, and secondary beam generation in fixed-target configurations \cite{Placzek:2019xpw,Curatolo:2018pza,GF-software-workshop-2021}. Together, these results provide strong evidence that the fundamental building blocks of the GF concept are technically viable.

The full implementation of the GF programme at CERN nevertheless requires the validation of several additional elements. These include: reliable and stable laser transport and injection into long Fabry-Perot cavities-- with performance comparable to state-of-the-art interferometric systems; fully remote and continuous operation of high-power laser systems in the demanding environment of accelerator tunnels; and precise, reproducible control of PSI beam parameters and of the spatial and temporal overlap between laser and ion beams. 

Further requirements include quantitative agreement between simulations and experimental measurements of atomic excitation and beam-cooling rates; the development of robust diagnostic tools for atomic and photon beams, including precise photon-flux measurements; the installation of PSI-specific collimation systems in the LHC; and the validation of schemes for extracting MW-class photon beams from the LHC ring and transporting them to experimental areas.

A distinctive strength of the GF approach is that the most critical challenges can be addressed through a dedicated, relatively low-cost Proof-of-Principle (PoP) experiment at the CERN SPS. Importantly, this experiment can be conducted in an environment more demanding in terms of radiation load and mechanical stability than that enticipated  for the final implementation in the LHC, thereby providing a conservative validation of the concept.

The GF PoP experiment \cite{GF-PoP-LoI:2019}, proposed in 2019, is designed to test all critical operational aspects of the programme. Its implementation benefits from the installation of the TT2 stripper system in the PS-to-SPS transfer line, which enables efficient production of Li-like Pb beam for the PoP experiment and permits parasitic  operation within the SPS cycle. This minimises interference with standard SPS operations supplying beams to the LHC and to the North Area experiments.

Progress toward the PoP experiment has been delayed by limited financial and personnel resources. However, key components, including the commercial laser system, have recently been procured and successfully tested with support from the Physics Beyond Colliders (PBC) funding scheme. Preparation of the Technical Design Report (TDR) is now underway, with completion targeted by the end of 2026.

Timely installation and operation of the GF PoP experiment before the start of the HL-LHC phase is a critical milestone. It would position the GF project for rapid deployment at the LHC, should strategic priorities shift toward innovative reuse of existing CERN infrastructure. 

In addition, formal institutional support is essential for securing external funding, in particular through programmes such as Horizon Europe, and for enabling the transition of the Gamma Factory from a validated concept to an operational research facility.

\section{Implementation aspects}

\vspace{5 mm}
{\it "The journey of a thousand miles begins with one step." -- Lao Tzu }
 \vspace{5 mm}

A key strength of the Gamma Factory (GF) programme is that it can be realised through targeted, incremental upgrades of the existing CERN accelerator infrastructure, rather than requiring the construction of a new facility. The required modifications are well identified and limited in scope. They include improvements to the SPS vacuum system; installation of PSI-beam-specific collimation in the warm sections of the LHC; construction of dedicated laser rooms adjacent to the accelerator tunnel; integration of Fabry--Perot resonators into selected sections of the collider rings; implementation of photon-beam extraction systems; and the creation of underground experimental caverns for the use of secondary and tertiary GF beams.

The GF can be deployed in a staged manner, ensuring minimal interference with the long-term operation of the CERN accelerator complex. In an initial phase (foreseen in the 2036--2041 timeframe), the programme could operate parasitically by using a small fraction of LHC running time for PSI beams, electron beams, and laser-cooled fully-stripped ion beams. This approach allows early scientific exploitation while preserving the primary LHC mission. 

A full extracted-beam research programme could begin after the completion of the baseline LHC physics programme, currently projected around 2041, or earlier if the discovery or precision reach of the LHC is saturated before that time. This staged deployment provides flexibility and reduces the risk associated with long-term planning.

The GF fixed-target programme could use two non-interacting, counter-propagating LHC beams composed of different atomic species with identical magnetic rigidity. This configuration enables efficient use of the existing machine layout allowing to produce simultaneously  both high- and low-energy photon beams. 

\begin{figure}[!htpb]\centering
\includegraphics[width=1.00\linewidth]{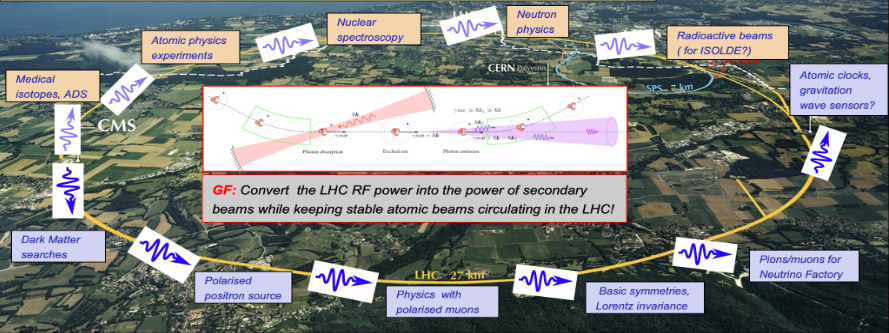}
\caption{Schematic layout of Gamma Factory photon-beam production and extraction points for selected physics applications. In the initial phase, a single laser station and experimental cavern would suffice to launch the programme. Additional stations can be implemented on different accelerator rings, provided that the PSI beams do not intersect.}
\label{Fig:photon_beams}
\end{figure}

The modular nature of the GF concept allows the progressive installation of laser interaction points and photon-beam extraction lines, each optimised for specific physics goals. Figure~\ref{Fig:photon_beams} illustrates possible configurations of laser stations and extraction points for a range of applications. In the early phase of the programme, a single laser station and a single experimental cavern would be sufficient to initiate the GF operations, with additional stations added as required.
The extracted photon beams can be distributed to multiple experimental setups, enabling parallel research activities. Beam collimation systems allow the selection of narrow-band photon beams tailored for high-precision spectroscopic measurements \cite{Budker:2021fts}. 

Overall, the implementation strategy of the Gamma Factory combines modularity, flexibility, and cost-effectiveness, allowing the programme to evolve progressively while delivering early scientific output and maintaining compatibility with CERN’s broader research agenda.

\section{Sustainability}

\vspace{5 mm}
{\it "There's a way to do it better -- find it." -- T. Edison}
 \vspace{5 mm}

The long-term sustainability of large-scale accelerator facilities is becoming an increasingly important consideration. CERN currently consumes of the order of 1.3~TWh of electrical energy per year, primarily to operate its accelerator complex. At 2025 market conditions, this corresponds to an annual cost of approximately 100~MCHF, representing a significant fraction of CERN’s operating budget. 

Future projects, such as the FCC-ee, will further increase energy demand and extend these costs over multiple decades. In this context, sustainability must be understood not only in environmental terms, but also in relation to economic stability and long-term operational resilience in a rapidly evolving and uncertain global energy landscape.

One of the central challenges is the increasing volatility of energy prices. Over the past five years, electricity prices have fluctuated by up to a factor six. While CERN has thus far managed such variations through long-term contracts and financial planning, it is uncertain whether this approach will remain sufficient over the multi-decade timescales relevant for future facilities. Price fluctuations are expected to intensify, driven both by short-term imbalances in renewable energy production and by longer-term instabilities in the supply and cost of conventional fuels.

As long as electricity remains reliably available at a manageable cost, continued reliance on the external power grid remains the most practical solution. However, a forward-looking strategy requires preparation for less favourable scenarios. In particular, it is prudent to assess whether CERN could, if necessary, secure a degree of energy autonomy through technologies that are independent of weather variability and external market fluctuations. Such a capability would not only enhance operational stability but could also strengthen public and institutional support by demonstrating a commitment to responsible and sustainable energy use.

In this context, the Gamma Factory (GF) programme has initiated exploratory studies of an Advanced Nuclear Energy System (ANES) \cite{Baolong:2024ataN}. This concept exploits the high-intensity GF photon beams to drive a compact, subcritical nuclear reactor. The proposed system could simultaneously generate thermal power at the level of several hundred megawatts—potentially sufficient to cover the energy needs of the CERN accelerator complex—and enable the transmutation of a significant fraction of long-lived radioactive waste.

\begin{figure}[!htpb]\centering
	\includegraphics[width=0.85\linewidth]{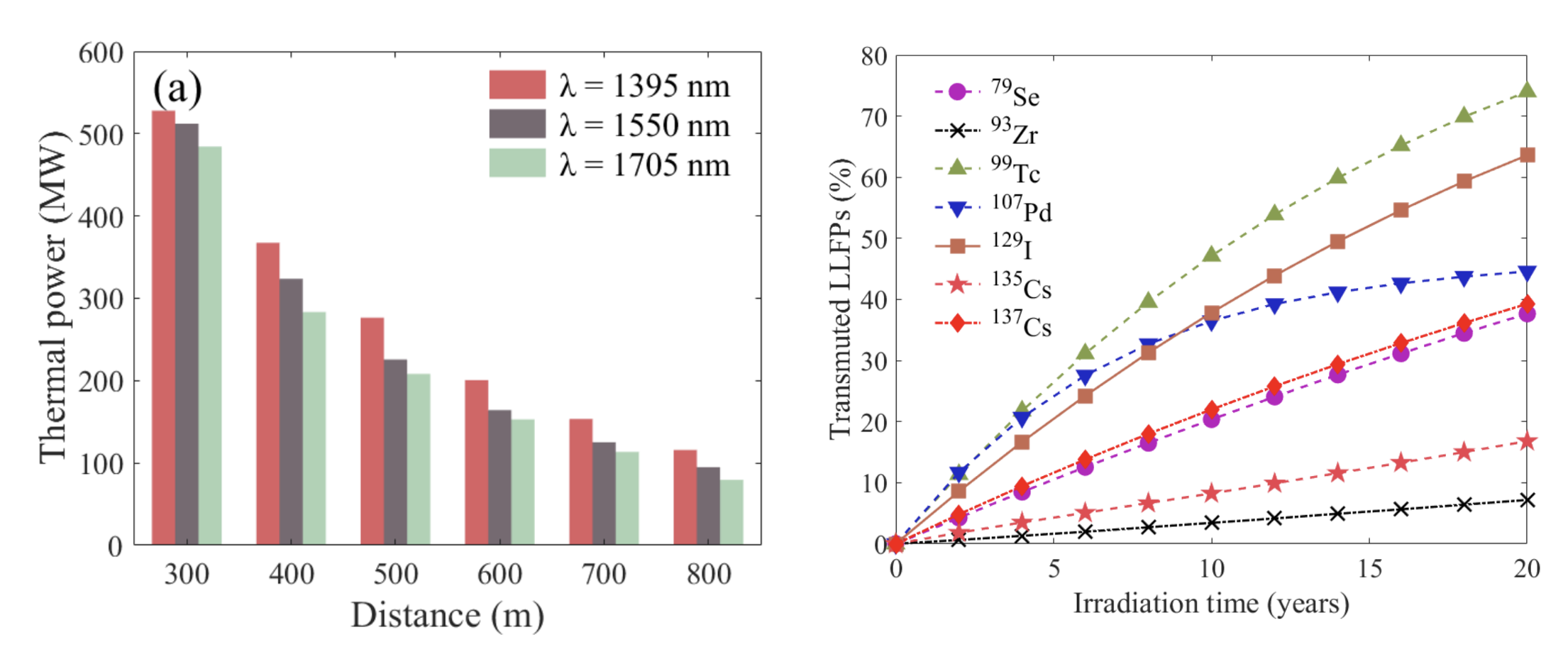}
	\caption{Left: thermal power of the Gamma-Factory-driven nuclear reactor for different laser-photon wavelengths; Right: transmutation capacity for long-lived fission products in the GF-driven ANES system. See \cite{Baolong:2024ataN} for details.}
    \label{fig:Power_transmutation}
\end{figure}

Even in scenarios where energy price volatility does not necessitate the development of an on-site power source, the GF-enabled photon-beam-driven transmutation programme would retain significant value. It would complement existing proton-driven transmutation studies by extending their reach to long-lived isotopes that can be efficiently processed through resonant photon-induced reactions \cite{Hayakawa2016_Transmut}. 

More broadly, such research would position CERN as a contributor not only to fundamental science but also to technologically relevant solutions addressing global challenges in energy production and nuclear waste management.

\section{The need for an HL-LHC--FCC research bridge}

\vspace{5 mm}
{\it "Do not fear to be eccentric in opinion, for every opinion now 
accepted was once eccentric." -- B. Russell}
\vspace{5 mm}

If a new high-energy frontier collider is constructed at CERN, 
it will most likely be the electron--positron Future Circular Collider 
(FCC-ee), as recommended in the European Strategy for Particle 
Physics \cite{ESPPU2026}. This option combines strong scientific 
motivation with a number of well-recognised practical advantages, 
which are worth enumerating explicitly.

The FCC-ee is supported by a large fraction of the world's particle 
physics community. It can be constructed using presently available 
technologies, and its detectors, computing infrastructure, and analysis 
frameworks can be developed through well-understood evolutionary steps. 
It ensures continuity in research methodology and collaboration 
structures, and can reuse established Monte Carlo simulation tools and 
the advanced theoretical frameworks elaborated over many decades. 
This continuity provides stability, reduces risk, and reinforces 
CERN's leading role in high-energy physics.

However, these same features also define the scope of the FCC-ee 
programme. It represents a continuation---albeit at a significantly 
enhanced level---of the current paradigm of collider-based particle 
physics. Complementary approaches that explore qualitatively different 
paradigms have received comparatively less attention in the European 
Strategy discussions.

The Gamma Factory (GF) project belongs to this latter category. 
While it shares with the FCC-ee the advantage of relying on existing 
technologies, it differs in several key respects. Its implementation 
would introduce new experimental methodologies, foster closer 
integration between the particle, nuclear, and atomic physics 
communities, and require the development of novel tools for handling 
quantum-state-controlled beams. Such a shift would naturally lead to 
a different research ecosystem, characterised by smaller, task-oriented 
teams and new interfaces between accelerator operation and 
experimentation.

In this context, it is not surprising that GF-like initiatives---despite 
their potential---have so far remained outside the main focus of the 
European Strategy process\footnote{The GF project is not explicitly 
named in the complementary experiments and non-collider priorities 
section of the Strategy document.}. Nevertheless, the question arises 
whether CERN's long-term planning should rely exclusively on a single, 
high-energy frontier trajectory, or whether complementary options 
should also be developed as part of a more robust and diversified 
strategy.

This question becomes particularly relevant in view of the uncertainties 
associated with large-scale, multi-decade infrastructure projects. 
The realisation timeline of the FCC-ee may be affected by a range of 
external factors, including geopolitical developments, funding 
constraints, and evolving international collaboration dynamics. Even 
under favourable conditions, a significant time gap between the end of 
the HL-LHC programme and the start of FCC-ee operations appears very 
likely.

Such a gap cannot be filled by injector-based fixed-target programmes 
alone, whose performance is already approaching its practical limits 
in terms of beam intensity and quality. This situation motivates the 
exploration of a complementary bridge programme capable of sustaining 
CERN's scientific activity and innovation potential during the 
transition period.

An effective HL-LHC--FCC bridge programme should satisfy the following 
conditions:
\begin{enumerate}
    \item It should open novel research opportunities uniquely enabled 
    by CERN's existing infrastructure.
    \item It should exploit existing facilities in innovative ways, 
    rather than requiring the construction of a new accelerator complex.
    \item Its cost should remain modest compared to that of a new 
    collider, at the level of O(100)~MCHF.
    \item It should be deployable on a timescale compatible with the 
    conclusion of the HL-LHC programme, so that it can begin operations 
    without delay.
    \item It should attract a broad scientific community, extending 
    beyond particle physics to include nuclear, atomic, and applied 
    physics, thereby strengthening the overall support base for CERN's 
    activities---particularly in the event that the HL-LHC programme 
    does not yield major particle physics discoveries.
    \item It should be compatible with long-term sustainability goals, 
    including resilience to energy cost fluctuations and supply 
    uncertainties.
\end{enumerate}

The Gamma Factory programme has the potential to satisfy all of these 
conditions. By enabling high-intensity, high-precision studies using 
the existing CERN accelerator infrastructure, it offers a complementary 
research direction that could remain scientifically productive 
throughout the transition period between major collider projects, while 
also opening qualitatively new avenues for discovery.

For the GF to fulfil this role, timely action is required. Explicit 
recognition of the project as a potential component of CERN's long-term 
strategy would be necessary, together with the implementation of its 
Proof-of-Principle experiment in the SPS during the current long 
shutdown period. These steps would allow the GF concept to mature to 
a level where it can be realistically considered as an operational 
programme.

It is important to emphasise that the Gamma Factory is not conceived 
as a competitor to the FCC-ee. On the contrary, it is a complementary 
initiative that could make productive use of the SPS and LHC 
infrastructure during a period when it would otherwise be 
underutilised, thereby enhancing the overall scientific return of 
CERN's accelerator complex.

\section{A lesson from the past?}

\vspace{5 mm}
{\it “History teaches us that man learns nothing from history.”  
— G. W. F. Hegel}
\vspace{5 mm}

The strategic choices faced by CERN today are not without precedent. 
A relevant example can be found in the evolution of the HERA programme at DESY in the 1990s.

Shortly after the start of HERA operations in 1992, it became clear that the facility could not compete with LEP and the Tevatron in precision electroweak measurements or in searches for new particles. 
However, early results quickly revealed an alternative and highly promising direction: the study of strong interactions using deep inelastic scattering. 
This opportunity motivated the proposal to extend the HERA programme to include electron--ion collisions \cite{EIC_at_HERA_Hera_worksop1,EIC_at_HERA_Seeheim_workshop_willeke,EIC_at_HERA_Seeheim_workshop,Krasny:1997mv}. 

The underlying idea was to use well-controlled electroweak probes to investigate the structure of strongly interacting matter across different regimes. 
By colliding electrons with protons and nuclei, one could study the transition between perturbative and non-perturbative QCD, using nuclei as femtoscopic analysers of the space--time structure of strong interactions. 
Such a programme would also have provided essential input for interpreting the heavy-ion collision data later collected at the LHC.

Between 1995 and 1999, a substantial international effort was devoted to developing the Electron--Ion Collider (EIC) concept at HERA \cite{EIC_at_HERA_HAmburg_1999_workshop}. 
Most importantly, this effort led to a convergence of several European initiatives, including ELFE, GSI-ENC, and HERA-based proposals. 
A realistic opportunity was created to establish at DESY a European centre dedicated to studies of strong interactions across a broad range of energy scales and distances.

In parallel, however, DESY was developing the TESLA electron--positron collider project, which was widely regarded as the primary strategic priority. 
In this context, the comparatively modest-cost HERA-based EIC programme—despite its strong scientific case—was effectively placed on hold. 
This decision was motivated in part by the desire to present TESLA as the single, focused future direction for the laboratory.

When the TESLA project was ultimately not approved, the opportunity to develop a European EIC programme had already been lost. The GSI had already started working on its FAIR project while European ELFE activities migrated to the TJNAF laboratory, and   
DESY-based  EIC studies were transferred to BNL   \cite{Krasny:1999av,EIC_at_BNL_Yale,EIC_at_BNL_Snowmass1,EIC_at_BNL_Snowmass2,Krasny:2001nin,EIC_at_BNL_first_white_paper,Krasny:2001tr}, contributing to the establishment of the 
present-day EIC project two decades later. The EIC project is now the leading accelerator construction effort in the United States. 

This historical episode suggests a broader lesson. 
A strategy focused exclusively on a single flagship project—however well justified—may overlook complementary opportunities that could provide scientific continuity and long-term resilience. 
In the case of DESY, the absence of a “safety net” option contributed to the discontinuity of its on-site particle physics programme and to the loss of leadership in a domain where it had a strong initial position.

The parallel with the present situation at CERN is not exact, but it is instructive. 
As discussed in this paper, the transition from the HL-LHC to a future collider such as the FCC-ee is likely to involve a significant temporal gap. 
Without complementary initiatives, this period risks being characterised by reduced experimental activity and missed opportunities for innovation.

Developing and supporting alternative programmes—such as the Gamma Factory—does not weaken the case for a future high-energy collider. 
On the contrary, it can strengthen the overall strategy by ensuring continuity of scientific output, fostering methodological diversity, and preserving flexibility in the face of evolving external conditions.

\section{Conclusions}

\vspace{5 mm}
{\it "The important thing is not to stop questioning." -- A. Einstein}
\vspace{5 mm}

The Gamma Factory project represents a rare convergence of scientific 
ambition, technological readiness, and strategic opportunity. 
Built upon the existing CERN accelerator infrastructure and 
state-of-the-art laser technology, it offers a path toward a 
qualitatively new experimental regime --- one that extends CERN's 
research programme far beyond its present boundaries without 
requiring the construction of a new major facility.
The scientific case for the GF is broad and compelling. 
It spans particle, nuclear, atomic, fundamental, and applied physics. 

The feasibility of the core GF concept has been demonstrated through 
a decade of studies. The production, acceleration, and storage of 
PSI beams at the SPS and LHC have been successfully validated. 
High-power Fabry--Perot cavities operating at record stored powers 
have been achieved. Dedicated simulation frameworks have been 
developed and benchmarked. The remaining steps toward full 
implementation are well identified and technically within reach, 
provided that adequate resources are allocated to the 
Proof-of-Principle experiment at the SPS.

From a strategic perspective, the GF addresses a genuine and 
urgent need. The transition from the HL-LHC era to a future 
high-energy frontier collider --- most likely the FCC-ee --- will 
involve a significant temporal gap that cannot be filled by 
injector-based programmes alone. The Gamma Factory offers a 
cost-effective, innovative, and scientifically rich bridge 
programme capable of sustaining CERN's experimental activity and 
its capacity for discovery throughout this transition period.

The historical precedent of the HERA programme at DESY illustrates 
the risks of a strategy that concentrates exclusively on a single 
flagship project. Complementary initiatives, if not actively 
supported, may lose critical momentum and ultimately migrate 
elsewhere --- to the detriment of the host laboratory and the 
broader European research community. The Gamma Factory represents 
precisely the kind of complementary initiative that merits 
protection from this outcome.

The GF is not a competitor to the FCC-ee. It occupies a different 
scientific niche, operates on a different timescale, and requires 
a fraction of the investment. Its implementation would strengthen, 
rather than weaken, CERN's long-term research strategy by 
introducing methodological diversity, broadening the scientific 
community engaged with CERN's facilities, and providing a degree 
of resilience against the uncertainties --- scientific, financial, 
and geopolitical --- that inevitably accompany multi-decade 
infrastructure planning.

Three actions are needed in the near term. First, the Gamma Factory 
should receive explicit recognition as a candidate component of 
CERN's long-term strategy, including its formal inclusion in the 
complementary experiments and non-collider priorities section of 
the European Strategy document. Second, the Proof-of-Principle 
experiment at the SPS should be installed and commissioned during 
the current long shutdown period, before the start of the HL-LHC 
phase. Third, institutional support should be secured to enable 
access to external funding mechanisms, in particular Horizon Europe, 
allowing the project to develop at the pace that its scientific 
and strategic potential warrants.

The Gamma Factory was conceived a decade ago.  
It has since grown into a broad, multidisciplinary 
programme supported by researchers across multiple fields and 
institutions. What it now requires is not a leap of faith, but a 
measured and timely institutional commitment --- the one which commensurates with the opportunity it represents.

\subsection*{Acknowledgements}
The development of the Gamma Factory project has been made possible 
by the dedicated work of a large number of enthusiastic researchers, 
whose efforts have continued despite limited recognition and 
institutional support from their home laboratories. Their names and 
affiliations are listed in the Appendix. I am deeply indebted to 
all of them.

\begin{flushleft}
{\LARGE\bf Appendix: \\ List of physicists who contributed to the Gamma Factory studies:}
\end{flushleft}
\noindent
A.~Abramov$^{1}$,
A.~Afanasev$^{37}$,
S.E.~Alden$^{1}$,
R.~Alemany~Fernandez$^{2}$,
P.S.~Antsiferov$^{3}$, 
A.~Apyan$^{4}$,
G.~Arduini$^{2}$,
D.~Balabanski$^{34}$,
B. ~Hu$^{41}$,
R.~Balkin$^{32}$
% ...


{\em
${}^{1}$ Royal Holloway University of London Egham, Surrey, TW20 0EX, United Kingdom\\
${}^{2}$ CERN, Geneva, Switzerland \\
${}^{3}$ Institute of Spectroscopy, Russian Academy of Sciences, Troitsk, Moscow Region, Russia\\
${}^{4}$ A.I.~Alikhanyan National Science Laboratory, Yerevan, Armenia\\
${}^{5}$ School of Physics, University of New South Wales, Sydney NSW 2052, Australia\\
${}^{6}$ P.N.~Lebedev Physical Institute, Russian Academy of Sciences, Moscow, Russia\\
${}^{7}$ Jagiellonian University, Krak\'ow, Poland \\
${}^{8}$ Center for Advanced Studies of Accelerators,  Jefferson Lab, USA\\
${}^{9}$ Helmholtz Institute,  Johannes Gutenberg University, Mainz, Germany\\
${}^{10}$ Department of Physics, University of California, Berkeley, CA 94720-7300, USA\\
${}^{11}$ Universit'{e} Paris Saclay, Laboratoire de Physique des 2 Infinis Irene Joliot Curie (IJCLab), Orsay, 
\\${}^{~~~}$
France
\\
${}^{12}$ Department of Physics, INFN--Milan  and University of Milan,  Milan, Italy\\
${}^{13}$ INFN--Padua,  Padua, Italy\\
${}^{14}$ University of Nevada, Reno, Nevada 89557, USA\\
${}^{15}$ University of Malta, Malta\\
${}^{16}$ LPNHE, University Paris Sorbonne, CNRS--IN2P3, Paris, France\\
${}^{17}$ HI Jena, IOQ FSU  Jena and GSI Darmstadt, Germany\\
${}^{18}$ Budker Institute of Nuclear Physics, Novosibirsk, Russia \\
${}^{19}$ Rochester Scientific, LLC, El Cerrito, CA 94530, USA\\
${}^{20}$ GSI Helmholtzzentrum f\"ur Schwerionenforschung, 64291 Darmstadt, Germany\\
${}^{21}$ Braunschweig University of Technology and Physikalisch-Technische Bundesanstalt, Germany \\
${}^{22}$ FEL Laboratory, Duke University,  Durham, USA\\
${}^{23}$ University of Padua, Padua, Italy\\
${}^{24}$ Center for Beam Physics, LBNL, Berkeley, USA\\
${}^{25}$ University of New South Wales, Sydney,  Australia\\
${}^{26}$ National Institutes for Quantum and Radiological Science and Technology (QST), Ibaraki, Japan \\ 
${}^{~~~}$ and Technology, Ibaraki, Japan\\
${}^{27}$ Institute of Modern Physics, Chinese Academy of Sciences, Lanzhou, China\\
${}^{28}$ Institut f\"ur Kernphysik, Johannes Gutenberg-Universit\"at, Mainz, Germany\\
${}^{29}$ Department of Physics and Astronomy, University of Delaware, Delaware, USA\\
${}^{30}$ Joint Quantum Institute, NIST and the University of Maryland, College Park, Maryland, USA\\
${}^{31}$ Laboratory of Advanced Science and Technology for Industry, University of Hyogo, Hyogo, Japan\\
${}^{32}$ Physics Department, Technion -- Israel Institute of Technology, Haifa 3200003, Israel\\
${}^{33}$ University of Science and Technology, Hefei (Anhui), China\\
${}^{34}$ Extreme Light Infrastructure -- Nuclear Physics (ELI-NP), 
Horia Hulubei National Institute for R\&D
\\${}^{~~~}$
in Physics and Nuclear Engineering (IFIN-HH), 077125 Bucharest-Magurele, Romania
\\
${}^{35}$ Laboratori Nazionali di Frascati dell'INFN, Via E. Fermi 54, Frascati (Roma), Italy
\\
${}^{36}$ Institut f\"ur Quantenoptik und Quanteninformation, Technikerstr.\ 21a, 6020 Innsbruck, Austria
\\
${}^{37}$ Department of Physics, George Washington University, Washington, DC 20052, USA
\\
${}^{38}$ Helmholtz-Zentrum Dresden-Rossendorf Dresden, Saxony, Germany 
\\
${}^{39}$ Helmholtz-Institut Jena, Fr\"obelstieg 3, 07743 Jena, Germany
\\
${}^{40}$ Laboratoire des Mat'{e}riaux Avances,
Institut de Physique des 2 Infinis de Lyon,
Campus LyonTech la Doua, batiment Virgo,
7 avenue Pierre de Coubertin, 69622 Villeurbanne
\\
${}^{41}$ South China University, Hengyang, China
\\
}



\end{document}